\newcommand\aj{{AJ\,}}%
\newcommand\apj{{ApJ\,}}%
\newcommand\apjl{{ApJ\,}}%
\newcommand\apjs{{ApJS\,}}%
\newcommand\aap{{A\&A\,}}%
\newcommand\mnras{{MNRAS\,}}%
\newcommand\nat{{Nature\,}}%
\documentclass[graybox, natbib, footinfo]{svmult}

\usepackage{mathptmx}       % selects Times Roman as basic font
\usepackage{helvet}         % selects Helvetica as sans-serif font
\usepackage{courier}        % selects Courier as typewriter font
\usepackage{type1cm}        % activate if the above 3 fonts are
\usepackage{makeidx}         % allows index generation
\usepackage{graphicx}        % standard LaTeX graphics tool
\usepackage{multicol}        % used for the two-column index
\usepackage[bottom]{footmisc}% places footnotes at page bottom

\usepackage{amssymb}

\makeindex             % used for the subject index
\begin{document}
\title*{Early Results from APOKASC}
% Use \titlerunning{Short Title} for an abbreviated version of
% your contribution title if the original one is too long
\author{Courtney Epstein}
% Use \authorrunning{Short Title} for an abbreviated version of
% your contribution title if the original one is too long
\institute{Courtney Epstein \at The Ohio State University, Department of Astronomy, McPherson Laboratory, 140 W 18th Avenue, Columbus, Ohio 43210-1173, \email{epstein@astronomy.ohio-state.edu}}
%
% Use the package "url.sty" to avoid
% problems with special characters
% used in your e-mail or web address
%
\maketitle

\abstract{Asteroseismology and spectroscopy provide complementary constraints on the fundamental and chemical properties of stars. I describe the first results from APOKASC, a collaboration between the \textit{Kepler} asteroseismic science consortium (KASC) and the SDSS-III APOGEE survey. These include (1) the first test of asteroseismic scaling relationships in the metal-poor regime using halo and thick disk stars identified in the APOKASC sample; and (2) the calibration of spectroscopic parameters using precise asteroseismic measurements of surface gravity. I also highlight future research avenues that are made possible by this unique sample of thousands of well-characterized red giant stars.}

\section{Introduction}

To understand galaxy formation and evolution, one can either study numerous high-redshift galaxies or examine in detail one nearby galaxy, like the Milky Way. Reconstructing the star formation history of the Milky Way requires detailed knowledge of the fundamental and chemical properties of its constituent stars. 

To achieve this goal, the Apache Point Observatory Galaxy Evolution Experiment (APOGEE; \cite{Majewski2010}) partnered with the  \textit{Kepler} asteroseismic science consortium (KASC) to form APOKASC. APOGEE is a high-resolution, high signal-to-noise, H-band spectroscopic survey that is part of the Sloan Digital Sky Survey III \citep{Eisenstein2011}. APOGEE provides measurements of effective temperature (T$_\mathrm{eff}$), surface gravity ($\log g$), metallicity ([M/H]), rotation ($v\sin i$), and radial velocity. In the \textit{Kepler} field, these spectroscopic measurements will be complemented by asteroseismic constraints on mass, radius, and $\log g$, and evolutionary state information, when available (e.g., \citealt{Bedding2011,Mosser2012,Stello2013}).

In the first year of observations \citep{Ahn2013}, APOGEE observed $\sim1,900$ red giants with asteroseismically determined parameters. These data will be made publicly available in the APOKASC catalog \cite{Pinsonneault2014}. I will describe two early results where the APOKASC dataset has been used to test the asteroseismic (\S\ref{sec:SR}) and spectroscopic (\S\ref{sec:logg}) determination of stellar parameters. I conclude by outlining future areas of investigation.

\section{Testing Asteroseismic Scaling Relationships}\label{sec:SR}

The APOKASC sample is useful for testing the asteroseismic mass scale. Two asteroseismic observables can be extracted from the power spectrum of oscillating red giants, namely the frequency of maximum power, $\nu_\mathrm{max}$, and the large frequency separation, $\Delta\nu$. The masses and radii of stars with oscillations driven by surface convection may be estimated using empirical scaling relationships \citep{Ulrich1986,Brown1991,Kjeldsen1995}:

\begin{eqnarray}
  \frac{\Delta \nu}{\Delta \nu_{\odot}} \simeq& \left(\frac{\mathrm{M}}{\mathrm{M}_{\odot}}\right)^{1/2} \left(\frac{\mathrm{R}}{\mathrm{R}_{\odot}}\right)^{-3/2}\label{eq:delta nu scaling relation} \\
  \frac{\nu_\mathrm{max}}{\nu_{\mathrm{max},\odot}} \simeq& \left(\frac{\mathrm{M}}{\mathrm{M}_{\odot}}\right)\left(\frac{\mathrm{R}}{\mathrm{R}_\odot}\right)^{-2}\left(\frac{\mathrm{T}_\mathrm{eff}}{\mathrm{T}_{\mathrm{eff,\odot}}}\right)^{-1/2}, \label{eq:nu max scaling relation}
\end{eqnarray}
where $\Delta \nu_\odot=135.0 \pm 0.1$ $\mu$Hz, $\nu_{\mathrm{max},\odot}=3140 \pm 30$ $\mu$Hz, and T$_{\mathrm{eff},\odot}=5777$ K \citep{Pinsonneault2014}.

These scaling relationships are calibrated based on the Sun and are therefore not guaranteed to work for more evolved stars, like red giants, which have a different internal structure. The accuracy of radii derived with asteroseismic scaling relationships have been examined using interferometry \citep{Huber2012}, \textit{Hipparcos} parallaxes \citep{SilvaAguirre2012}, and RGB stars in the open cluster NGC6791 \citep{Miglio2012} (see \citealt{Miglio2013_Differential_Population_Studies} for a review of constraints). 

Eclipsing binaries provide a test of scaling relationships masses. In particular, the metal-rich ([Fe/H]$\sim+0.4$ dex) open cluster NGC 6791 serves as a prime test case. From measurements of eclipsing binaries near the cluster's main sequence turn-off, the mass of red giant branch stars is inferred to be M$_\mathrm{RGB} = 1.15 \pm 0.02$ M$_\odot$ \citep{Brogaard2012}. Asteroseismic oscillations have been detected for red giant branch stars in this cluster. Scaling relationships yield mass estimates of M$_\mathrm{RGB}=1.20 \pm 0.01$ M$_\odot$ and $1.23 \pm 0.02$ M$_\odot$  from \cite{Basu2011} and \cite{Miglio2012}. These mass estimates are sensitive to, for example, the choice of temperature scale. \citet{Wu2014} defined an additional scaling relationship and determined a mass of M$_{\rm RGB} = 1.24 \pm 0.03$ M$_\odot$, consistent with previous results.

\subsection{Current Results: Probing the Metal-Poor Regime}

These tests of the standard scaling relationships have been confined to metallicities close to solar ($-0.5\lesssim\mathrm{[Fe/H]}\lesssim +0.4$). However, the scaling relationships take no account of metallicity. Because metallicity could potentially influence mode excitation and damping and opacity-driven changes in convective properties, stellar properties derived using scaling relationships need to be verified over a wide range of metallicities. The APOKASC sample enabled the first test of mass estimates from asteroseismic scaling relationships in the low-metallicity regime \citep{Epstein2014}.

In the first year of APOGEE observations in the \textit{Kepler} field, there were nine stars with measured asteroseismic parameters, reliable spectroscopic measurements, and [M/H]$<-1$ dex. We differentiated between halo and disk stars by computing the 3-D space velocities and adopting $v_{\rm tot,LSR}=180$ km/s as the kinematical division between stars classified as halo and disk populations \citep{Venn2004}. We considered metal-poor stars with disk kinematics to be members of the thick disk.

\begin{figure}[tb]
\sidecaption
\includegraphics[width=4.9in]{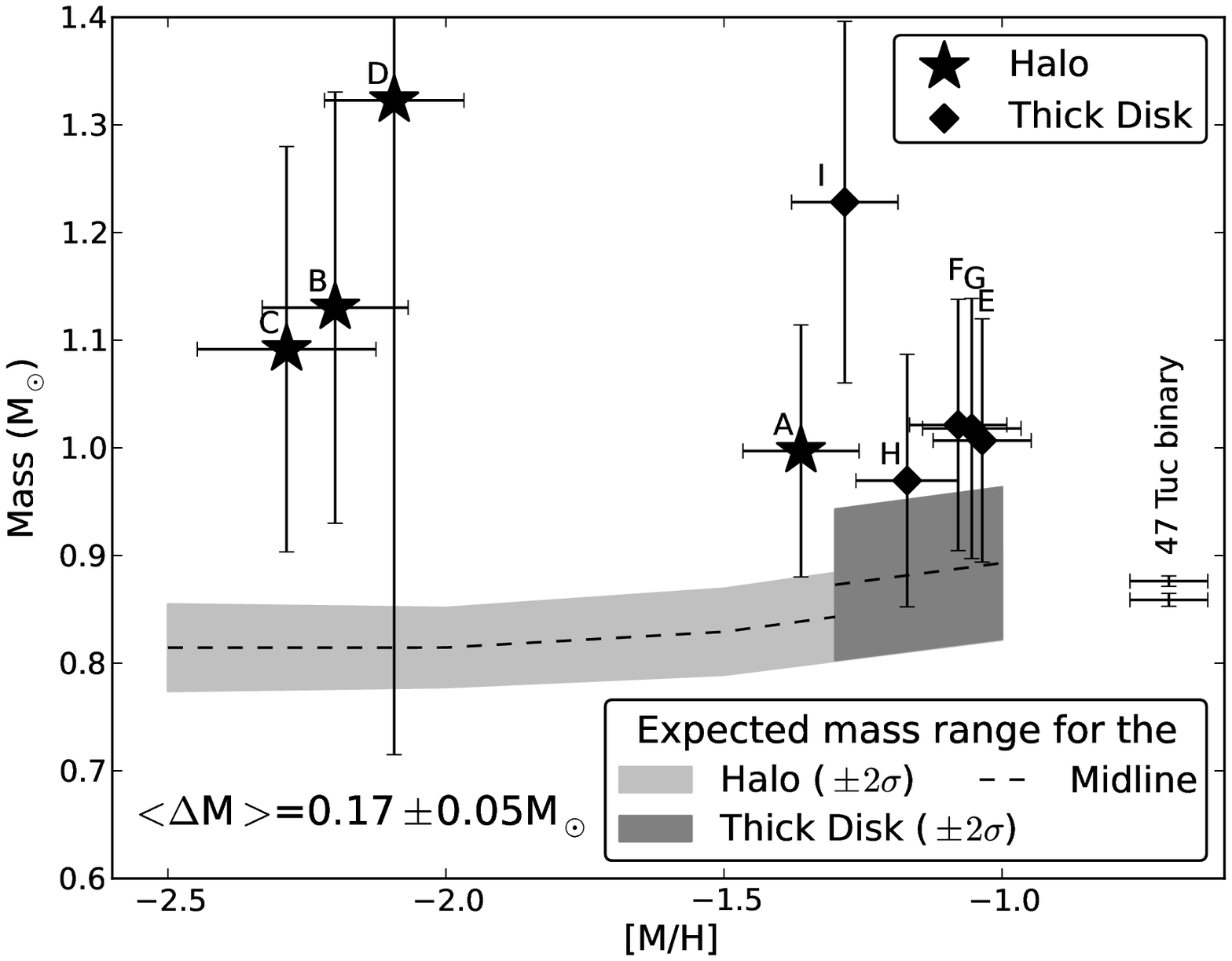}
\caption{Mass determined using standard asteroseismic scaling relationships (see Equations \ref{eq:delta nu scaling relation} and \ref{eq:nu max scaling relation}), asteroseismic parameters from the APOKASC Catalog \cite{Pinsonneault2014}, and spectroscopic temperatures from APOGEE. The expected mass ranges based on astrophysical priors for halo and thick disk stars are indicated by the light and dark gray bands, respectively. See \cite{Pinsonneault2014} for a description of the method used to determine $\Delta \nu$ and $\nu_\mathrm{max}$ and their uncertainties.}
\label{fig:mass-metallicity}
\end{figure}

Halo and metal-poor thick disk stars have a variety of independent constraints on their mass and age. The combined evidence from isochrone fits to globular clusters, e.g., \citet{Gratton1997}, the white dwarf cooling sequence, e.g., \citet{Hansen2002}, the radioactive decay of uranium and thorium, e.g., \citet{Sneden1996}, and the imprint left by the MSTO on the stellar temperature distribution function \citep{Jofre2011}, places halo stars as being 10 Gyr or older. Metal-poor, $\alpha$-enhanced members of the Galactic disk have been found to be 8 Gyr or older \citep{Bensby2013_Disk,Haywood2013}. We converted these age constraints into mass constraints using Dartmouth stellar isochrones \citep{Dotter2008}. The mass range associated with these age constraints is shown as the gray bands in Figure \ref{fig:mass-metallicity}. These mass ranges are consistent with the mass measurements of the eclipsing binary found in the metal-poor thick disk globular cluster 47 Tuc \citep{Thompson2010}. The width of these bands includes uncertainties in the input physics, including the assumed helium abundance, heavy element mixture, nuclear reaction rates, equation of state, opacity, model atmosphere, and heavy element diffusion rate (see \citealt{vanSaders2012} for details).

Figure \ref{fig:mass-metallicity} shows that the masses determined using asteroseismic scaling relationships, asteroseismic parameters from the APOKASC catalog, and spectroscopic temperatures are a weighted average of $\Delta$M$=0.17\pm0.05$ M$_\odot$ higher than expectations. Possible explanations for this mass discrepancy are fully detailed in \cite{Epstein2014}. For example, shifts in the temperature scale and contamination of this red giant branch sample by more evolved (e.g., horizontal branch or asymptotic giant branch) stars work to increase the mass difference. Two published theoretically motivated corrections \citep{White2011_Temperature_Correction,Mosser2013} reduce the mass estimates by as much as $\sim5\%$. Differences between methods of determining $\Delta \nu$ and $\nu_\mathrm{max}$ can shift $\Delta$M by as much as $0.04$ M$_\odot$ for these stars. Epstein et al.\ also note a difference between masses derived using scaling relationships compared with $\nu_\mathrm{max}$-independent techniques \citep{Bedding2006,Deheuvels2012}.

\subsection{Upcoming Work}

The second year of APOGEE observations will include additional metal-poor stars, enabling a better statistical test of the mass offset at low-metallicity. Detailed frequency modeling of metal-poor stars in the APOKASC sample will help to calibrate masses derived using scaling relationships. The influence of Equation \ref{eq:nu max scaling relation} on asteroseismic mass estimates also merits investigation. Theoretical corrections, e.g.,  \cite{White2011_Temperature_Correction}, have been computed for $-0.2<$[Fe/H]$<+0.2$; work is underway to extend similar models to lower metallicity.

\section{Calibrating Stellar Parameters from Atmospheric Models}\label{sec:logg}

Many studies have investigated the accuracy of asteroseismic gravities for evolved stars and have found good agreement between asteroseismic and spectroscopic techniques (e.g., \citealt{Morel2012,Thygesen2012}). Asteroseismic $\log g$ have been shown to be largely model independent \citep{Gai2011} and robust to using different approaches to derive oscillation parameters and incorporating corrections to the $\Delta\nu$ scaling relationship (Equation \ref{eq:delta nu scaling relation}) \citep{Hekker2013}.

The APOGEE survey used the asteroseismic $\log g$ as an independent check on the $\log g$ derived using the APOGEE Stellar Parameters and Chemical Abundances Pipeline (ASPCAP) \citep{Meszaros2013}. The ASPCAP determines stellar parameters (including $\log g$) and abundances by performing a $\chi^2$ minimization in a grid of synthetic spectra. M\'{e}sz\'{a}ros et al.\ defined an empirical correction to the ASPCAP $\log g$ based on combined constraints from asteroseismology and open/globular cluster stars for application to the larger APOGEE sample \citep{Meszaros2013}. 

\section{Future Applications and Conclusions}

\begin{figure}[tb]
\sidecaption
\includegraphics[width=4.7in]{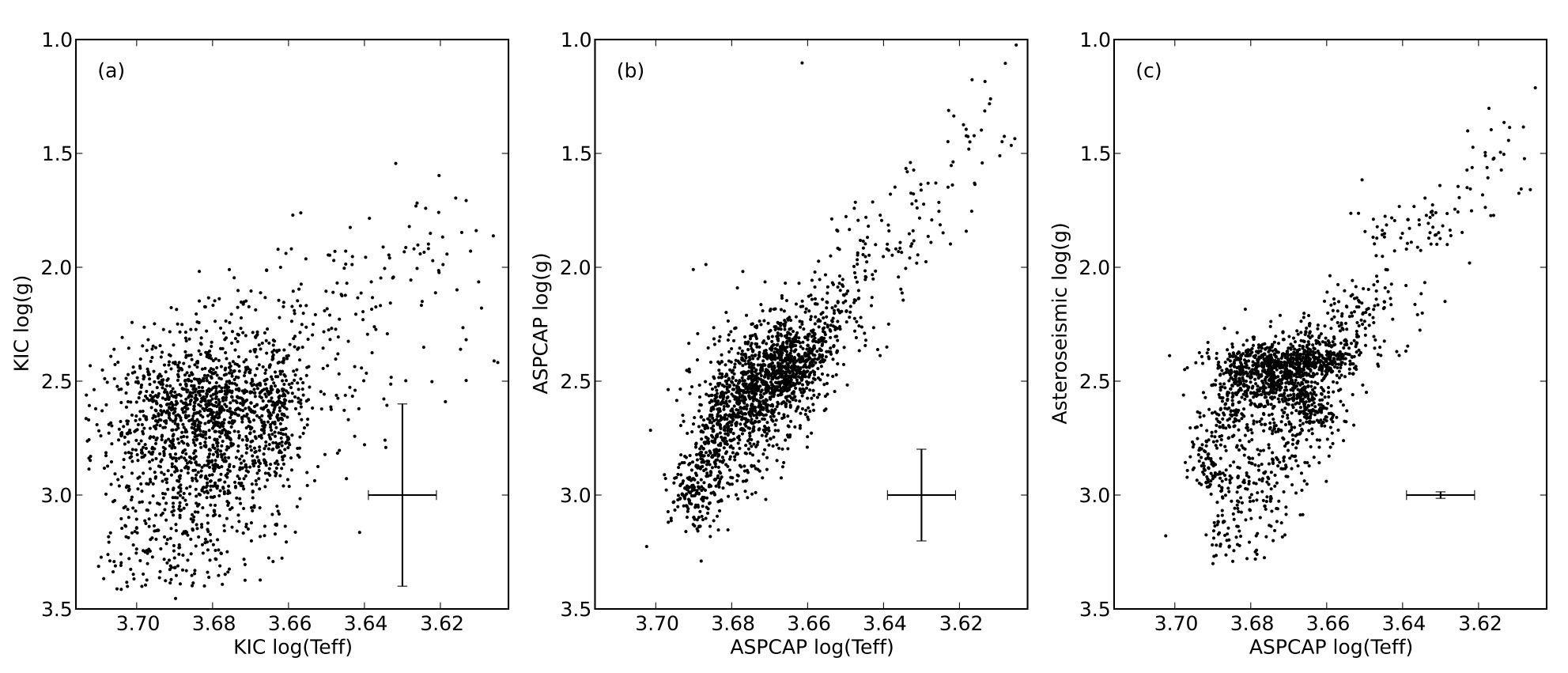}
\caption{A comparison of the ASPCAP sample's $\log g$ and T$_\mathrm{eff}$ distribution on the red giant branch, as determined using different measurement techniques. \textit{(a)} Parameters were taken from the Kepler Input Catalog, which is based on photometric inputs \cite{Brown2011}. {(b)} Spectroscopic parameters were derived using the ASPCAP. {(c)} Spectroscopic T$_\mathrm{eff}$ and asteroseismic $\log g$ from the APOKASC Catalog \cite{Pinsonneault2014}. The typical uncertainty is shown in the lower right.}
\label{fig:ohsnap}
\end{figure}

The APOKASC sample represents the largest set of joint asteroseismic and spectroscopic constraints available to date. Figure \ref{fig:ohsnap} shows the impact of these improved constraints on the H-R diagram. The precise asteroseismic $\log g$ measurements make possible the identification of features, like the red clump and red bump, in a sample of field stars. 

The detection of mixed-modes \cite{Bedding2011,Mosser2012,Stello2013} can distinguish between helium-core-burning and hydrogen-shell-burning stars. This information will be useful for identifying stellar populations and probing their properties as a function of a Galactic position and age. Bayesian estimates of stellar age (e.g., \citealt{Serenelli2013}) based on APOKASC's and asteroseismic constraints will help to reconstruct the star formation history of the Milky Way through diagnostics, such as the age-metallicity relationship.

Additionally, the locus of the red giant branch will provide an excellent test of stellar interior models. For example, we can test whether adopting a solar mixing-length for stars is consistent with asteroseismic measurements across a range of mass, [M/H], and $\log g$ (e.g., \citealt{Bonaca2012}). The power of precise asteroseismic $\log g$ (\S\ref{sec:logg}) could also be harnessed to constrain stellar model atmospheres.

The APOKASC collaboration offers a treasure trove of information that will enable studies of both stellar physics and stellar populations for years to come.

%%\input{referenc}
%% BibTeX users please use
%\bibliographystyle{spphys}
%\bibliography{mybib}

\begin{thebibliography}{40}
\expandafter\ifx\csname natexlab\endcsname\relax\def\natexlab#1{#1}\fi

\bibitem[{{Ahn} {et~al.}(2014){Ahn}, {Alexandroff}, {Allende Prieto}, {Anders},
  {Anderson}, {Anderton}, {Andrews}, {Aubourg}, {Bailey}, {Bastien}, \&
  et~al.}]{Ahn2013}
{Ahn}, C.~P., {Alexandroff}, R., {Allende Prieto}, C., {et~al.} 2014, \apjs,
  211, 17

\bibitem[{{Basu} {et~al.}(2011){Basu}, {Grundahl}, {Stello}, {Kallinger},
  {Hekker}, {Mosser}, {Garc{\'{\i}}a}, {Mathur}, {Brogaard}, {Bruntt},
  {Chaplin}, {Gai}, {Elsworth}, {Esch}, {Ballot}, {Bedding}, {Gruberbauer},
  {Huber}, {Miglio}, {Yildiz}, {Kjeldsen}, {Christensen-Dalsgaard},
  {Gilliland}, {Fanelli}, {Ibrahim}, \& {Smith}}]{Basu2011}
{Basu}, S., {Grundahl}, F., {Stello}, D., {et~al.} 2011, \apjl, 729, L10

\bibitem[{{Bedding} {et~al.}(2006){Bedding}, {Butler}, {Carrier}, {Bouchy},
  {Brewer}, {Eggenberger}, {Grundahl}, {Kjeldsen}, {McCarthy}, {Nielsen},
  {Retter}, \& {Tinney}}]{Bedding2006}
{Bedding}, T.~R., {Butler}, R.~P., {Carrier}, F., {et~al.} 2006, \apj, 647, 558

\bibitem[{{Bedding} {et~al.}(2011){Bedding}, {Mosser}, {Huber},
  {Montalb{\'a}n}, {Beck}, {Christensen-Dalsgaard}, {Elsworth},
  {Garc{\'{\i}}a}, {Miglio}, {Stello}, {White}, {De Ridder}, {Hekker}, {Aerts},
  {Barban}, {Belkacem}, {Broomhall}, {Brown}, {Buzasi}, {Carrier}, {Chaplin},
  {di Mauro}, {Dupret}, {Frandsen}, {Gilliland}, {Goupil}, {Jenkins},
  {Kallinger}, {Kawaler}, {Kjeldsen}, {Mathur}, {Noels}, {Aguirre}, \&
  {Ventura}}]{Bedding2011}
{Bedding}, T.~R., {Mosser}, B., {Huber}, D., {et~al.} 2011, \nat, 471, 608

\bibitem[{{Bensby} {et~al.}(2013){Bensby}, {Feltzing}, \&
  {Oey}}]{Bensby2013_Disk}
{Bensby}, T., {Feltzing}, S., \& {Oey}, M.~S. 2013, ArXiv e-prints

\bibitem[{{Bonaca} {et~al.}(2012){Bonaca}, {Tanner}, {Basu}, {Chaplin},
  {Metcalfe}, {Monteiro}, {Ballot}, {Bedding}, {Bonanno}, {Broomhall},
  {Bruntt}, {Campante}, {Christensen-Dalsgaard}, {Corsaro}, {Elsworth},
  {Garc{\'{\i}}a}, {Hekker}, {Karoff}, {Kjeldsen}, {Mathur}, {R{\'e}gulo},
  {Roxburgh}, {Stello}, {Trampedach}, {Barclay}, {Burke}, \&
  {Caldwell}}]{Bonaca2012}
{Bonaca}, A., {Tanner}, J.~D., {Basu}, S., {et~al.} 2012, \apjl, 755, L12

\bibitem[{{Brogaard} {et~al.}(2012){Brogaard}, {VandenBerg}, {Bruntt},
  {Grundahl}, {Frandsen}, {Bedin}, {Milone}, {Dotter}, {Feiden}, {Stetson},
  {Sandquist}, {Miglio}, {Stello}, \& {Jessen-Hansen}}]{Brogaard2012}
{Brogaard}, K., {VandenBerg}, D.~A., {Bruntt}, H., {et~al.} 2012, \aap, 543,
  A106

\bibitem[{{Brown} {et~al.}(1991){Brown}, {Gilliland}, {Noyes}, \&
  {Ramsey}}]{Brown1991}
{Brown}, T.~M., {Gilliland}, R.~L., {Noyes}, R.~W., \& {Ramsey}, L.~W. 1991,
  \apj, 368, 599

\bibitem[{{Brown} {et~al.}(2011){Brown}, {Latham}, {Everett}, \&
  {Esquerdo}}]{Brown2011}
{Brown}, T.~M., {Latham}, D.~W., {Everett}, M.~E., \& {Esquerdo}, G.~A. 2011,
  \aj, 142, 112

\bibitem[{{Deheuvels} {et~al.}(2012){Deheuvels}, {Garc{\'{\i}}a}, {Chaplin},
  {Basu}, {Antia}, {Appourchaux}, {Benomar}, {Davies}, {Elsworth}, {Gizon},
  {Goupil}, {Reese}, {Regulo}, {Schou}, {Stahn}, {Casagrande},
  {Christensen-Dalsgaard}, {Fischer}, {Hekker}, {Kjeldsen}, {Mathur}, {Mosser},
  {Pinsonneault}, {Valenti}, {Christiansen}, {Kinemuchi}, \&
  {Mullally}}]{Deheuvels2012}
{Deheuvels}, S., {Garc{\'{\i}}a}, R.~A., {Chaplin}, W.~J., {et~al.} 2012, \apj,
  756, 19

\bibitem[{{Dotter} {et~al.}(2008){Dotter}, {Chaboyer}, {Jevremovi{\'c}},
  {Kostov}, {Baron}, \& {Ferguson}}]{Dotter2008}
{Dotter}, A., {Chaboyer}, B., {Jevremovi{\'c}}, D., {et~al.} 2008, \apjs, 178,
  89

\bibitem[{{Eisenstein} {et~al.}(2011){Eisenstein}, {Weinberg}, {Agol},
  {Aihara}, {Allende Prieto}, {Anderson}, {Arns}, {Aubourg}, {Bailey},
  {Balbinot}, \& et~al.}]{Eisenstein2011}
{Eisenstein}, D.~J., {Weinberg}, D.~H., {Agol}, E., {et~al.} 2011, \aj, 142, 72

\bibitem[{{Epstein} {et~al.}(2014){Epstein}, {Elsworth}, {Johnson}, {Shetrone},
  {Mosser}, {Hekker}, {Tayar}, {Harding}, {Pinsonneault}, {Silva Aguirre},
  {Basu}, {Beers}, {Bizyaev}, {Bedding}, {Chaplin}, {Frinchaboy},
  {Garc{\'{\i}}a}, {Garc{\'{\i}}a P{\'e}rez}, {Hearty}, {Huber}, {Ivans},
  {Majewski}, {Mathur}, {Nidever}, {Serenelli}, {Schiavon}, {Schneider},
  {Sch{\"o}nrich}, {Sobeck}, {Stassun}, {Stello}, \& {Zasowski}}]{Epstein2014}
{Epstein}, C.~R., {Elsworth}, Y.~P., {Johnson}, J.~A., {et~al.} 2014, \apjl,
  785, L28

\bibitem[{{Gai} {et~al.}(2011){Gai}, {Basu}, {Chaplin}, \&
  {Elsworth}}]{Gai2011}
{Gai}, N., {Basu}, S., {Chaplin}, W.~J., \& {Elsworth}, Y. 2011, \apj, 730, 63

\bibitem[{{Gratton} {et~al.}(1997){Gratton}, {Fusi Pecci}, {Carretta},
  {Clementini}, {Corsi}, \& {Lattanzi}}]{Gratton1997}
{Gratton}, R.~G., {Fusi Pecci}, F., {Carretta}, E., {et~al.} 1997, \apj, 491,
  749

\bibitem[{{Hansen} {et~al.}(2002){Hansen}, {Brewer}, {Fahlman}, {Gibson},
  {Ibata}, {Limongi}, {Rich}, {Richer}, {Shara}, \& {Stetson}}]{Hansen2002}
{Hansen}, B.~M.~S., {Brewer}, J., {Fahlman}, G.~G., {et~al.} 2002, \apjl, 574,
  L155

\bibitem[{{Haywood} {et~al.}(2013){Haywood}, {Di Matteo}, {Lehnert}, {Katz}, \&
  {Gomez}}]{Haywood2013}
{Haywood}, M., {Di Matteo}, P., {Lehnert}, M., {Katz}, D., \& {Gomez}, A. 2013,
  ArXiv e-prints

\bibitem[{{Hekker} {et~al.}(2013){Hekker}, {Elsworth}, {Mosser}, {Kallinger},
  {Basu}, {Chaplin}, \& {Stello}}]{Hekker2013}
{Hekker}, S., {Elsworth}, Y., {Mosser}, B., {et~al.} 2013, \aap, 556, A59

\bibitem[{{Huber} {et~al.}(2012){Huber}, {Ireland}, {Bedding}, {Brand{\~a}o},
  {Piau}, {Maestro}, {White}, {Bruntt}, {Casagrande}, {Molenda-{\.Z}akowicz},
  {Silva Aguirre}, {Sousa}, {Barclay}, {Burke}, {Chaplin},
  {Christensen-Dalsgaard}, {Cunha}, {De Ridder}, {Farrington}, {Frasca},
  {Garc{\'{\i}}a}, {Gilliland}, {Goldfinger}, {Hekker}, {Kawaler}, {Kjeldsen},
  {McAlister}, {Metcalfe}, {Miglio}, {Monteiro}, {Pinsonneault}, {Schaefer},
  {Stello}, {Stumpe}, {Sturmann}, {Sturmann}, {ten Brummelaar}, {Thompson},
  {Turner}, \& {Uytterhoeven}}]{Huber2012}
{Huber}, D., {Ireland}, M.~J., {Bedding}, T.~R., {et~al.} 2012, \apj, 760, 32

\bibitem[{{Jofr{\'e}} \& {Weiss}(2011)}]{Jofre2011}
{Jofr{\'e}}, P. \& {Weiss}, A. 2011, \aap, 533, A59

\bibitem[{{Kjeldsen} \& {Bedding}(1995)}]{Kjeldsen1995}
{Kjeldsen}, H. \& {Bedding}, T.~R. 1995, \aap, 293, 87

\bibitem[{{Majewski} {et~al.}(2010){Majewski}, {Wilson}, {Hearty}, {Schiavon},
  \& {Skrutskie}}]{Majewski2010}
{Majewski}, S.~R., {Wilson}, J.~C., {Hearty}, F., {Schiavon}, R.~R., \&
  {Skrutskie}, M.~F. 2010, in IAU Symposium, Vol. 265, IAU Symposium, ed.
  K.~{Cunha}, M.~{Spite}, \& B.~{Barbuy}, 480--481

\bibitem[{{M{\'e}sz{\'a}ros} {et~al.}(2013){M{\'e}sz{\'a}ros}, {Holtzman},
  {Garc{\'{\i}}a P{\'e}rez}, {Allende Prieto}, {Schiavon}, {Basu}, {Bizyaev},
  {Chaplin}, {Chojnowski}, {Cunha}, {Elsworth}, {Epstein}, {Frinchaboy},
  {Garc{\'{\i}}a}, {Hearty}, {Hekker}, {Johnson}, {Kallinger}, {Koesterke},
  {Majewski}, {Martell}, {Nidever}, {Pinsonneault}, {O'Connell}, {Shetrone},
  {Smith}, {Wilson}, \& {Zasowski}}]{Meszaros2013}
{M{\'e}sz{\'a}ros}, S., {Holtzman}, J., {Garc{\'{\i}}a P{\'e}rez}, A.~E.,
  {et~al.} 2013, \aj, 146, 133

\bibitem[{{Miglio} {et~al.}(2012){Miglio}, {Brogaard}, {Stello}, {Chaplin},
  {D'Antona}, {Montalb{\'a}n}, {Basu}, {Bressan}, {Grundahl}, {Pinsonneault},
  {Serenelli}, {Elsworth}, {Hekker}, {Kallinger}, {Mosser}, {Ventura},
  {Bonanno}, {Noels}, {Silva Aguirre}, {Szabo}, {Li}, {McCauliff}, {Middour},
  \& {Kjeldsen}}]{Miglio2012}
{Miglio}, A., {Brogaard}, K., {Stello}, D., {et~al.} 2012, \mnras, 419, 2077

\bibitem[{{Miglio} {et~al.}(2013){Miglio}, {Chiappini}, {Morel}, {Barbieri},
  {Chaplin}, {Girardi}, {Montalb{\'a}n}, {Noels}, {Valentini}, {Mosser},
  {Baudin}, {Casagrande}, {Fossati}, {Silva Aguirre}, \&
  {Baglin}}]{Miglio2013_Differential_Population_Studies}
{Miglio}, A., {Chiappini}, C., {Morel}, T., {et~al.} 2013, in European Physical
  Journal Web of Conferences, Vol.~43, European Physical Journal Web of
  Conferences, 3004

\bibitem[{{Morel} \& {Miglio}(2012)}]{Morel2012}
{Morel}, T. \& {Miglio}, A. 2012, \mnras, 419, L34

\bibitem[{{Mosser} {et~al.}(2012){Mosser}, {Goupil}, {Belkacem}, {Michel},
  {Stello}, {Marques}, {Elsworth}, {Barban}, {Beck}, {Bedding}, {De Ridder},
  {Garc{\'{\i}}a}, {Hekker}, {Kallinger}, {Samadi}, {Stumpe}, {Barclay}, \&
  {Burke}}]{Mosser2012}
{Mosser}, B., {Goupil}, M.~J., {Belkacem}, K., {et~al.} 2012, \aap, 540, A143

\bibitem[{{Mosser} {et~al.}(2013){Mosser}, {Michel}, {Belkacem}, {Goupil},
  {Baglin}, {Barban}, {Provost}, {Samadi}, {Auvergne}, \&
  {Catala}}]{Mosser2013}
{Mosser}, B., {Michel}, E., {Belkacem}, K., {et~al.} 2013, \aap, 550, A126

\bibitem[{{Pinsonneault} {et~al.}(in prep.)}]{Pinsonneault2014}
{Pinsonneault}, M.~H. {et~al.} in prep.

\bibitem[{{Serenelli} {et~al.}(2013){Serenelli}, {Bergemann}, {Ruchti}, \&
  {Casagrande}}]{Serenelli2013}
{Serenelli}, A.~M., {Bergemann}, M., {Ruchti}, G., \& {Casagrande}, L. 2013,
  \mnras, 429, 3645

\bibitem[{{Silva Aguirre} {et~al.}(2012){Silva Aguirre}, {Casagrande}, {Basu},
  {Campante}, {Chaplin}, {Huber}, {Miglio}, {Serenelli}, {Ballot}, {Bedding},
  {Christensen-Dalsgaard}, {Creevey}, {Elsworth}, {Garc{\'{\i}}a}, {Gilliland},
  {Hekker}, {Kjeldsen}, {Mathur}, {Metcalfe}, {Monteiro}, {Mosser},
  {Pinsonneault}, {Stello}, {Weiss}, {Tenenbaum}, {Twicken}, \&
  {Uddin}}]{SilvaAguirre2012}
{Silva Aguirre}, V., {Casagrande}, L., {Basu}, S., {et~al.} 2012, \apj, 757, 99

\bibitem[{{Sneden} {et~al.}(1996){Sneden}, {McWilliam}, {Preston}, {Cowan},
  {Burris}, \& {Armosky}}]{Sneden1996}
{Sneden}, C., {McWilliam}, A., {Preston}, G.~W., {et~al.} 1996, \apj, 467, 819

\bibitem[{{Stello} {et~al.}(2013){Stello}, {Huber}, {Bedding}, {Benomar},
  {Bildsten}, {Elsworth}, {Gilliland}, {Mosser}, {Paxton}, \&
  {White}}]{Stello2013}
{Stello}, D., {Huber}, D., {Bedding}, T.~R., {et~al.} 2013, \apjl, 765, L41

\bibitem[{{Thompson} {et~al.}(2010){Thompson}, {Kaluzny}, {Rucinski},
  {Krzeminski}, {Pych}, {Dotter}, \& {Burley}}]{Thompson2010}
{Thompson}, I.~B., {Kaluzny}, J., {Rucinski}, S.~M., {et~al.} 2010, \aj, 139,
  329

\bibitem[{{Thygesen} {et~al.}(2012){Thygesen}, {Frandsen}, {Bruntt},
  {Kallinger}, {Andersen}, {Elsworth}, {Hekker}, {Karoff}, {Stello},
  {Brogaard}, {Burke}, {Caldwell}, \& {Christiansen}}]{Thygesen2012}
{Thygesen}, A.~O., {Frandsen}, S., {Bruntt}, H., {et~al.} 2012, \aap, 543, A160

\bibitem[{{Ulrich}(1986)}]{Ulrich1986}
{Ulrich}, R.~K. 1986, \apjl, 306, L37

\bibitem[{{van Saders} \& {Pinsonneault}(2012)}]{vanSaders2012}
{van Saders}, J.~L. \& {Pinsonneault}, M.~H. 2012, \apj, 746, 16

\bibitem[{{Venn} {et~al.}(2004){Venn}, {Irwin}, {Shetrone}, {Tout}, {Hill}, \&
  {Tolstoy}}]{Venn2004}
{Venn}, K.~A., {Irwin}, M., {Shetrone}, M.~D., {et~al.} 2004, \aj, 128, 1177

\bibitem[{{White} {et~al.}(2011){White}, {Bedding}, {Stello},
  {Christensen-Dalsgaard}, {Huber}, \&
  {Kjeldsen}}]{White2011_Temperature_Correction}
{White}, T.~R., {Bedding}, T.~R., {Stello}, D., {et~al.} 2011, \apj, 743, 161

\bibitem[{{Wu} {et~al.}(2014){Wu}, {Li}, \& {Hekker}}]{Wu2014}
{Wu}, T., {Li}, Y., \& {Hekker}, S. 2014, \apj, 781, 44

\end{thebibliography}

% Mac users: please ignore the error message: "! Package natbib Error: Bibliography not compatible with author-year citations."
\end{document}